\documentclass[journal]{IEEEtran}

\usepackage{amsmath}
\usepackage{amsfonts}
\usepackage{amssymb}
\usepackage[]{graphicx}
\usepackage{bm}
\usepackage{subfigure}
\usepackage{ragged2e}
\usepackage{setspace}
\usepackage{color}

\ifCLASSINFOpdf
\else
\fi

\begin{document}
%
\title{A Statistical Framework for Optimizing and Evaluating MRI $T_1$ and $T_2$ Relaxometry Approaches}
%
%
%

\author{Yang~Liu, John R. Buck and Shaokuan Zheng
            \thanks{Dr. Yang Liu is with the Consumer Electronics Division, Bose Corporation, 100 the Mountain Rd, Framingham, MA  01701 USA (e-mail: yangliu$\_$acoustics@outlook.com). Dr. John R . Buck is with the Department of Electrical and Computer Engineering, University of Massachusetts Dartmouth, 285 Old Westport Rd, Dartmouth,  MA 02747 USA (e-mail: jbuck@umassd.edu). Dr. Shaokuan Zheng is with the Department of Radiology, University of Massachusetts Medical School, Worcester, MA 01655 USA (EMAIL: Shaokuan.Zheng@umassmed.edu)}}

\maketitle

\begin{abstract}
This paper proposes a statistical framework to optimize and evaluate the MR parameter $T_1$ and $T_2$ mapping capabilities for quantitative MRI relaxometry approaches. This analysis explores the intrinsic MR parameter estimate precision per unit scan time, termed the $T_{1,2}$-to-noise ratio (TNR) efficiency, for different ranges of biologically realistic relaxation times. The TNR efficiency is defined in terms of the Cramer-Rao bound (CRB), a statistical lower bound on the parameter estimate variance. Geometrically interpreting the new TNR efficiency definition reveals a more complete model describing the factors controlling the $T_1$/$T_2$ mapping capabilities. This paper compares $T_1$ mapping approaches including the inversion recovery (IR) family sequences and the Look-Locker (LL) sequence and simultaneous $T_1$ and $T_2$ mapping approaches including the spin-echo inversion recovery (SEIR) and driven equilibrium single pulse observation of $T_1$/$T_2$ (DESPOT) sequences. All pulse parameters are optimized to maximize the TNR efficiency within different $T_1$ and $T_2$ ranges of interest. Monte Carlo simulations with non-linear least square estimation (NLSE) of $T_1$/$T_2$ validated the theoretical predictions on the estimator performances.

\end{abstract}

\begin{IEEEkeywords}
quantitative MRI, relaxometry, $T_1$/$T_2$ mapping, efficiency, Cramer-Rao lower bound.
\end{IEEEkeywords}

%
\IEEEpeerreviewmaketitle

\section{Introduction}
Quantitative MRI estimates pixel-wise maps of MR quantities such as longitudinal relaxation time $T_1$ and transverse relaxation time $T_2$. These maps prove clinically useful in anatomical and functional studies as well as in neurodegenerative pathology evaluations. Quantitative maps provide more objective diagnostic information compared against the more commonly used qualitative images weighted by a subjective blend of the MR quantities.  

Literature \cite{IR}-\cite{despot} etc. have contributed to MR relaxation parameter estimation in the past three decades. Magnetization perturbation based pulse sequences such as inversion recovery (IR) have been the mainstream methods for $T_1$ estimation due to their relatively broad signal dynamic ranges \cite{IR}-\cite{FIR}. To reduce the lengthy scan time required by the IR sequences, Look and Locker proposed a `one-shot' Look-Locker (LL) sequence which samples the $T_1$ relaxation curve following the inversion pulse by repeatedly tipping the magnetization to the transverse plane through small flip angle pulses \cite{LL}. To estimate $T_1$ and $T_2$ simultaneously from a single experiment set-up rather than from two or more different ones, Kleef etc. proposed one sequence consisting of a multiple spin-echo (SE) sequence interleaved with a multiecho IR sequence for simultaneous $T_1$ and $T_2$ mapping \cite{RLSQ}. Alternatively, $T_1$ and $T_2$ information can be extracted simultaneously from steady state imaging methods. Clinically popular spoiled gradient echo (SPGR) imaging \cite{SPGR} and steady state free precession (SSFP) imaging \cite{SSFP} were adapted and re-named as driven equilibrium single pulse observation of $T_1$/$T_2$ (DESPOT) for fast $T_1$ and $T_2$ mapping \cite{despot}.

Despite the extensive relaxometry protocol design research for quantitative MRI, little effort has been made to statistically quantify the $T_1$/$T_2$ mapping efficiencies for different relaxometry approaches. The goal of this paper is to set up a solid mathematical framework for evaluating and optimizing the intrinsic $T_1$/$T_2$ estimate precision per unit time of different relaxometry approaches. This framework considers relaxometry protocols as estimation algorithms and proposes two new metrics: $T_1$/$T_2$-to-noise ratio (TNR) to characterize $T_1/T_2$ estimates' precisions and TNR efficiency to measure $T_1/T_2$ estimates' precision per unit time. Prior studies \cite{despot}-\cite{oneshot} discussed the MR parameter mapping efficiencies, however, these comparison metrics are limited in the way that they either 1) define the efficiency metric through an `ad-hoc' routine following the noise propagation theorem of least squares fit, 2) fail to include all of the factors impacting the estimates' precisions or 3) optimize the pulse sequence parameters while targeting at only specific nominal $T_1$ and $T_2$ values rather than a range of realistic target tissue values. 

To explore the intrinsic performances of different relaxometry approaches in $T_1$ and $T_2$ mappings, this paper defines the TNR efficiency using the Cramer-Rao bound (CRB), a statistical lower bound on the parameter estimation variance \cite{VanTrees}. The CRB approach has been used as a quantitative tool for designing experimental MR protocols \cite{DAlexander}\cite{TSICRB} and evaluating the precision of specific relaxometry methods \cite{Funai}-\cite{LiuISMRM}. The CRB allows us to predict the best case performance of an MRI pulse sequence based on parameters such as pulse timings and flip angles. Moreover, this bound allows us to optimize the performance of that sequence by adjusting these parameters. In this paper, the CRB is interpreted geometrically \cite{CRBgeometric} and for the first time utilized to define the $T_1$/$T_2$ estimate efficiency and applied in a systematic manner to compare the MR parameter mapping capabilities of different relaxometry approaches. Before comparison, the pulse parameters of each relaxometry approach are optimized for pre-selected ranges of relevant $T_1$ and $T_2$ values, corresponding to human brain white matter (WM) and grey matter (GM). The TNR efficiency defined in this paper can be used to reliably predict the $T_1/T_2$ mapping performances of any relaxometry approaches before any phantom or in vivo experiments.

\section{Theory} 
\subsection{Signal Model}
\label{SignalTheory}
For relaxometry based on either magnetization perturbation or the steady-state method, the acquired magnetic signal $\textbf{y}_{N \times 1}$ at all acquisition times $t_1,...,t_N$ is modeled by
\begin{equation}
\label{signalmodel}
\textbf{y} = \textbf{s}+\textbf{n} = M_0 \textbf{h}(T_1,T_2,...;\boldsymbol\gamma)+ \textbf{n},
\end{equation}
where $M_0$ is the equilibrium longitudinal magnetization. The signal weighting vector $\textbf{h}_{N \times 1}$ describes the signals' dependence on the tissue characteristics ($T_1$, $T_2$, magnetization transfer, etc.) and the specific pulse sequence parameters $\boldsymbol\gamma$ (pulse timings, flip angles, etc.). Vector $\textbf{n}_{N \times 1}$ is the additive noise at acquisition time $t_1,...,t_N$. For simplicity, the signal model in (\ref{signalmodel}) assumes monoexponential relaxation behavior for the magnetization system. The noise in magnitude MR images is generally modeled as Rician distribution. However, when the acquired signal amplitude is more than two times the noise floor, the Rician distribution converges to Gaussian with mean and variance related to the mean and variance of the Rician distribution \cite{Rician}. Therefore, the noise term $\textbf{n}$ in (1) is additive white and Gaussian with zero mean and known variance $\sigma^2$. This assumption simplifies the derivation of the analytic CRB expressions. The derived CRB can be geometrically interpreted to understand the competing factors controlling the $T_1$/$T_2$ mapping capabilities. 

For illustrative purpose, this paper considers several representational MRI relaxometry sequences. Specifically, the $T_1$ relaxometry sequences include the inversion-recovery (IR) based sequences \cite{IR}\cite{SR}\cite{FIR} and the Look-Locker (LL) sequence \cite{LL}\cite{MRISequenceBook}. The joint $T_1$/$T_2$ relaxometry sequences include the spin-echo inversion recovery (SEIR) sequences \cite{RLSQ} and the DESPOT sequences \cite{despot}. In the interest of brevity, we do not explicitly list out the signal model for each relaxometry sequence, which are well presented in the given references. However, we do clarify the several IR-family sequences this paper considers due to their different forms and applications. In general, the IR sequence measures $T_1$ by varying the inversion time TI and the magnetic signal acquired immediately after the $90^o$ pulse follows \cite{IR}
\begin{equation}
\label{IR}
S_{\text{IR}} = M_0 [1- (2-e^{-W/T_1})e^{-\text{TI}/T_1}],
\end{equation}   
where $W$ is the wait time between successive measurements. The conventional inversion-recovery (CIR) sequence keeps the delay time $W$ constant and constrains $W\geq5T_1$ while varying TI for $T_1$ estimation and therefore, (\ref{IR}) simplifies to
\begin{equation}
\label{CIR}
S_{\text{CIR}} = M_0 (1-2e^{-\text{TI}/T_1}).
\end{equation}
The saturation recovery (SR) sequence keeps only the $90^o$ pulses and the acquired signals follow \cite{SR}
\begin{equation}
\label{SR}
S_{\text{SR}} = M_0(1-e^{-\text{TI}/T_1}).
\end{equation}
The fast inversion recovery (FIR) sequence varies the CIR sequence by allowing the wait time $W<5T_1$ to reduce the total scan time and the signal model follows (\ref{IR}). The FIR approach estimates $T_1$ by varying $\text{TI}$ while either fixing the wait time W (FIR1) or fixing the sequence repetition time TR (FIR2) \cite{FIR}. This paper considers both FIR approaches because both of them have practical implementations on different MRI scanners. 

\subsection{CRB for Simultaneous MR Parameter Estimation}
\label{CRBsection}
This section presents the analytic CRB expressions for the simultaneous estimation of the MR parameters $\boldsymbol \theta = [M_0, T_{1,2}]$, where the notation $T_{1,2}$ represents the $T_1$ and $T_2$ in the sequel. For the signal model in (1), define $\hat{\boldsymbol \theta}$ as the unbiased estimate of $\boldsymbol \theta$ from the noisy measurements $\textbf{y}$. Note that $M_0$ is included in the parameter vector $\boldsymbol\theta$ due to the acquired signals' dependence on $M_0$, although the spin density is less useful than $T_1$ and $T_2$ maps in providing quantitative information \cite{Liang}. The estimator covariance matrix $\bf{C(\hat{\boldsymbol \theta})}$ quantifies the precision of an estimator. The covariance matrix satisfies the inequality
\begin{equation}
\label{CRLBEq}
\bf{C(\hat{\theta})}-\bf{I^{-1}(\boldsymbol \theta)}\geq\bf{0},
\end{equation}
where $\bf{I(\boldsymbol \theta)}$ is the Fisher information matrix (FIM) \cite{Estimation}. Consequently, the diagonal entries of $\bf{I^{-1}(\theta)}$ lower bound the variances of any unbiased estimation of $\boldsymbol \theta$, known as the CRB of $\boldsymbol \theta$
\begin{equation}
\nonumber
\text{Var}(\hat{M}_0) \geq \left(\textbf{I}^{-1}(\boldsymbol \theta)\right)_{11} = \text{CRB}(M_0),
\end{equation}
\begin{equation}
\nonumber
\text{Var}(\hat{T}_1) \geq \left(\textbf{I}^{-1}(\boldsymbol \theta)\right)_{22} = \text{CRB}(T_1),
\end{equation}
\begin{equation}
\text{Var}(\hat{T}_2) \geq \left(\textbf{I}^{-1}(\boldsymbol \theta)\right)_{33} = \text{CRB}(T_2).
\end{equation} The FIM $\bf{I(\boldsymbol\theta)}$ in (\ref{CRLBEq}) captures the signals' sensitivities to $\boldsymbol \theta$ in noisy measurements, as well as the correlations between the parameters. Calculating the FIM is straightforward 
\begin{equation}
\textbf{I}(\boldsymbol\theta)=E\left([\nabla_{\boldsymbol\theta} \ln p(\textbf{y};\boldsymbol\theta)][\nabla_{\boldsymbol\theta} \ln p(\textbf{y};\boldsymbol\theta)]^T\right),
\end{equation}
where $E(\cdot)$ is the expectation operator and $\nabla$ is the derivative operator. Therefore, the FIM $\bf{I(\boldsymbol\theta)}$ is populated by the expected curvatures of $\boldsymbol\theta$'s likelihood function $p(\textbf{y};\boldsymbol \theta)$. For the case of simultaneous estimation of $\boldsymbol \theta = [M_0, T_{1,2}]$ in white Gaussian noise, the $3\times3$ FIM can be simplified as \cite{CRBmcDESPOT}
\begin{equation}
\label{simplifiedFIM}
\textbf{I}(\boldsymbol\theta) = \textbf{J}^T\textbf{J}/\sigma^2,
\end{equation}
where $\sigma^2$ is the noise variance (assumed uniform across all acquired signals) and $\textbf{J}_{N\times3}$ is a Jacobian matrix with the $i^{th}$ column $\textbf{J}_{N \times i} = \partial \textbf{s}/\partial \boldsymbol \theta_i$.  
This paper focuses on $T_1$ and $T_2$ maps since these maps are more clinically useful than $M_0$ maps \cite{Liang}. However, $M_0$ is included as a nuisance parameter in all analyses and simulations of simultaneous estimation procedures. The analytic CRB expressions are derived and interpreted geometrically in a linear space. For both $T_1$ relaxometry approaches and simultaneous $T_1$/$T_2$ relaxometry approaches, the CRB of $T_1$ and $T_2$ follow the same form of expression
\begin{equation}
\label{CRBT}
\text{CRB}(T_{1,2})= (\text{SNR} \cdot \text{Sens} \cdot \text{Orth})^{-2},
\end{equation}
where $\text{SNR} = M_0/\sigma$ and 
\begin{equation}
\label{Sens}
\text{Sens} = \begin{cases} \left| \left|\frac{\partial \textbf{h}}{\partial T_{1}}\right|\right| & \mbox{for } T_1 ~\text{relaxometry}, \\ 
~ \\
\left| \left|\frac{\partial \textbf{h}}{\partial T_{j}}\right|\right|, j = 1,2 & \mbox{for } T_{1,2} ~\text{relaxometry}, \end{cases}
\end{equation}
and 
\begin{equation}
\nonumber
\text{Orth} = ~~~~~~~~~~~~~~~~~~~~~~~~~~~~~~~~~~~~~~~~~~~~~~~~~~~~~~~~
\end{equation}
\vspace{-1em}
\begin{equation}
\label{Orth}
\begin{cases} ~~~~~~~~~~~~~\sin\phi_{1} & \mbox{for } T_1 ~\text{relaxometry}, \\ 
~\\
\frac{\sqrt{1+2\prod\limits_{i=1}^3 \cos \phi_i - \sum\limits_{i=1}^3 \cos^2 \phi_i}}{\sin \phi_{3-j}}, j = 1,2 & \mbox{for } T_{1,2} ~\text{relaxometry}.  \end{cases}
\end{equation}

In Eqs. (\ref{Sens}-\ref{Orth}), the signal weighting vector $\textbf{h}$ follows the definition in (1) and $\phi_1$ describes the geometric angle\footnote{Mathematically, the geometric angle $\phi$ between two vectors $x$ and $y$ in the linear space is defined as $\phi = \cos^{-1}\frac{\langle x,y\rangle}{\Vert x \Vert \cdot \Vert y \Vert}$. The $\langle\cdot\rangle$ is the inner product operator and $\Vert \cdot \Vert$ calculates the Euclidean norm (or length) of a vector.} between vector $\textbf{h}$ and $\partial\textbf{h}/\partial{T_1}$ in the linear space; $\phi_2$ describes the geometric angle between $\textbf{h}$ and $\partial\textbf{h}/\partial{T_2}$; $\phi_3$ describes the geometric angle between $\partial\textbf{h}/\partial{T_1}$ and $\partial\textbf{h}/\partial{T_2}$. Eq. (\ref{CRBT}) reveals that the CRB on $T_{1,2}$ depend on three characteristics of the acquired MR signals: the SNR, the signals' sensitivity to $T_{1,2}$ and the orthogonality between three vectors $\textbf{h},~\partial\textbf{h}/\partial{T_1}$ and $\partial\textbf{h}/\partial{T_2}$. 
\subsection{TNR Efficiency Definition}
\label{SNREfficiency}
The MR parameter mapping performance is measured by how precisely and accurately a relaxometry approach estimates the MR quantities given a fixed total scan time. To this end, we define a performance metric in terms of the precision per unit time. This metric describes the efficiency of a relaxometry approach in generating a $T_1$ or $T_2$ map. The precision of the resulting parameter map can be alternatively treated as the output $\text{SNR}$ of the parameter estimation process, that is, the true parameter value $T_{1,2}$ over the standard deviation of the parameter estimate $\sigma(\hat{T}_{1,2})$. To avoid confusion with the input $\text{SNR} = M_0/\sigma$, we define a new term $T_{1,2}$-to-noise ratio (TNR) to represent the output SNR for $T_1$/$T_2$ estimation
\begin{equation}
\text{TNR} = \frac{T_{1,2}}{\sigma(\hat{T}_{1,2})}.
\end{equation} 
Assuming the noise is uncorrelated between signal acquisitions, the TNR of $T_{1,2}$ estimation improves for $N_{\text{rep}}$ signal acquisitions due to signal averaging
\begin{equation}
\text{TNR} = \frac{T_{1,2}}{\sigma(\hat{T}_{1,2})}\sqrt{N_{\text{rep}}}.
\end{equation}
In 2D Cartesian DFT image reconstruction, each pixel signal is averaged $N_{\text{rep}} = N_{\text{ex}}\cdot N_{\text{PE}}$ times \cite{oneshot}, where $N_{\text{ex}}$ is the number of sequence repetitions and $N_{\text{PE}}$ is the number of phase encodings. When each $T_{1,2}$ estimate requires scan time $T_{\text{seq}}$, the total scan time $T_{\text{scan}} = N_{\text{rep}}\cdot T_{\text{seq}}$. Therefore the TNR changes into
\begin{equation}
\text{TNR} = \frac{T_{1,2}}{\sigma(\hat{T}_{1,2})}\sqrt{\frac{T_{\text{scan}}}{T_{\text{seq}}}}.
\end{equation}
The TNR efficiency is defined as the TNR per unit scan time
\begin{equation}
\label{Gamma}
\Gamma_{1,2} = \frac{T_{1,2}}{\sigma(\hat{T}_{1,2})\cdot\sqrt{T_{\text{seq}}}}.
\end{equation}

As discussed in the previous section, the $\text{CRB}(T_{1,2})$ lower bounds the parameter estimation variance $\text{Var}(\hat{T}_{1,2})$. Therefore, the maximum possible TNR efficiency results from replacing $\sigma (\hat{T}_{1,2})$ in (\ref{Gamma}) with $\sqrt{\text{CRB}(T_{1,2})}$
\begin{equation}
\label{efficiency}
\Gamma_{1,2} = \frac{T_{1,2}}{\sqrt{\text{CRB}(T_{1,2})}\cdot\sqrt{T_{\text{seq}}}} = \frac{T_{1,2} \cdot \text{SNR} \cdot \text{Sens} \cdot \text{Orth}}{\sqrt{T_{\text{seq}}}},
\end{equation}
where the input $\text{SNR}$, the signal sensitivity and the orthogonality follow (\ref{CRBT}-\ref{Orth}). Note that the sequence time $T_{\text{seq}}$ is different from the pulse sequence repetition time TR, although these two quantities are related due to the multiple sequence repetitions required for MR parameter estimation. The specific relations between $T_{\text{seq}}$ and TR for each relaxometry approach considered in this paper are given in Table \ref{table1}. 

Eq. (\ref{efficiency}) indicates that for a fixed total scan time, decreasing the sequence time $T_{\text{seq}}$ improves the TNR efficiency through signal averaging. The TNR efficiency depends linearly on the input $\text{SNR} = M_0/\sigma$ and the signal sensitivity $||\partial \textbf{h}/\partial T_{1,2}||$. Prior studies \cite{despot}-\cite{oneshot} included either one or both of these two terms in defining the MR parameter mapping efficiencies. However, Eq. (\ref{efficiency}) indicates that increasing the orthogonality term also improves the TNR efficiency without any further demands on the hardware or scan time. This is the first time the signal orthogonality is incorporated in the definition of the efficiency for MR parameter mapping. Therefore, we claim that the CRB provides a more complete model for the factors controlling MR relaxometry performance efficiency.

\section{Methods}
\label{SeqOptimization}

Before comparing the performances of the relaxometry approaches mentioned in the theory section, the sequence parameters of each approach are optimized to maximize the TNR efficiency in (\ref{efficiency}). Realistically, the proton density $M_0$, $T_1$, $T_2$ vary in heterogeneous biological tissues. Therefore, the optimization algorithm targets a range of $T_1$/$T_2$ values corresponding to the tissue of interest in optimizing the pulse parameters, rather than assuming any nominal $T_1$/$T_2$ value.
 
The optimization process uses the max-min criterion to achieve overall optimality of the sequence parameters within specified $T_1$/$T_2$ ranges. The optimization cost function is defined as a weighted sum of the $T_1$ estimate efficiency $\Gamma_1$ and $T_2$ estimate efficiency $\Gamma_2$
\begin{equation}
\label{costfunction}
\Lambda = \rho\Gamma_1+(1-\rho)\Gamma_2,
\end{equation}
where $0\leq\rho\leq1$ is the weighting coefficient for $T_1$ estimate efficiency. For $T_1$ only relaxometry, $\rho = 1$. Note that the cost function $\Lambda$ depends on the sequence parameters including the pulse timings $\textbf{t} = [t_1,...,t_N,\text{TR}]$, pulse flip angles $\boldsymbol{\alpha} = [\alpha_1,...,\alpha_N]$ and also the unknown parameters to be estimated $\boldsymbol{\theta} = [M_0, T_{1,2}]$. The max-min criterion works by first finding the values of $\boldsymbol\theta$ over the $T_1$/$T_2$ ranges of interest that gives the minimum (which is the worst) case of the weighted TNR efficiency $\Lambda$ for a given set of pulse parameters. The worst case efficiency $\Lambda_{\text{min}}$ is a function of the pulse parameters only
\begin{equation}
\Lambda_{\text{min}}(\textbf{t},\boldsymbol\alpha) = \min_{\boldsymbol\theta_{\text{min}}\leq \boldsymbol \theta \leq \boldsymbol\theta_{\text{max}}} \Lambda(\textbf{t},\boldsymbol\alpha,\boldsymbol\theta).
\end{equation}
The optimal sequence parameters are then chosen as those maximizing the worst case efficiency $\Lambda_{\text{min}}(\text{t},\boldsymbol \alpha)$ while satisfying protocol constraints on pulse sequence timings and angles: 
 \begin{eqnarray}
\label{constraint}
(\textbf{t},\boldsymbol\alpha)^{\text{opt}} & = & \arg ~ \max_{\textbf{t},\boldsymbol \alpha} ~\Lambda_{\text{min}}(\textbf{t},\boldsymbol\alpha), \\
s.t. ~~~ \textbf{t},\boldsymbol\alpha & \in & C,     
\nonumber
\end{eqnarray}where $C$ represents the specific pulse constraints. To test the optimization approach, this paper considers one particular set of $T_1 \in [1000, 2000]$ ms and $T_2\in [60, 110]$ ms, corresponding to human brain white and grey matters values measured \textit{in vitro} at 3.0T \cite{ReferenceT1T2}. The sampling times for each sequence assume to be linear spaced in time along the $T_1$ or $T_2$ relaxation curve \cite{oneshot}. Multiple runs of the optimization algorithm with different initial pulse parameter choices are seeded to aid finding relatively global optimal solutions.

For all Monte Carlo simulations, the IR-family and LL sequences simultaneously estimate $M_0$ and $T_1$ and the SEIR and DESPOT sequences simultaneously estimate $M_0$, $T_1$ and $T_2$. Non-linear least square estimation (NLSE) is used to estimate all unknown parameters due to its optimality in the sense of achieving the CRB if the bound is attainable \cite{Poor}. All estimated $T_1$ and $T_2$ values from Monte Carlo simulations are recorded to calculate the statistical mean and variance of each relaxometry approach. The numerical values of the variance validate the theoretical predictions on $T_1$/$T_2$ estimate precisions. This validation proves whether the CRB is a tight bound and furthermore the validity of the TNR efficiency definition for each relaxometry approach. For example, at a $T_1 = 1500$ ms in 10 seconds of total scan time, a TNR efficiency of 17 converts to a precision of $\pm 27.9$ ms (or $\pm$ 1.86$\%$) for $T_1$ estimation. Statistically, the precision is measured by the mean error of the estimation (MEE), defined as the standard deviation of the estimated values over the true $T_{1,2}$ value
\begin{equation}
\label{mse}
\text{MEE}(\hat{T}_{1,2}) = \frac{\sigma(\hat{T}_{1,2})}{T_{1,2}}.
\end{equation}
The smaller the MEE, the more precise the estimation. The accuracy is measured by the relative bias (Rbias) of the estimator, defined as the difference between the mean of the estimated values $\hat{T}_{1,2}$ and the true $T_{1,2}$, normalized over the true $T_{1,2}$ value
\begin{equation}
\label{PBias}
\text{Rbias}(\hat{T}_{1,2}) = \frac{E(\hat{T}_{1,2})-T_{1,2}}{T_{1,2}}.
\end{equation}
Positive Rbias implies over estimation and negative Rbias implies under estimation. The smaller the absolute value of the relative bias, the more accurate the estimation.




\section{Results}
\subsection{Sequence Optimization Results}

\begin{table*}[!t]

\caption{Optimized pulse sequence parameters for different relaxometry approaches with $T_1 = [1000, 2000]$ ms and $T_2 = [60, 110]$ ms. Note that for notation $t = [a:b:c]$, $a$ is the start value, $b$ is the step value and $c$ is the end value. N is the number of elements in vector t. All sequence parameter optimizations and TNR efficiency calculations assume an input SNR $M_0/\sigma = 100$.}
\begin{center}

\begin{tabular}{|c|c|c|c|}
\hline 
Approach & Optimized sequence parameters (ms) & $T_{\text{seq}}$ (ms) & TNR Efficiency \\ 
\hline 
DESPOT & $\alpha_{\text{SPGR}} = 8.6^o$, $\alpha_{\text{SSFP}} = [13.9^o, 57.8^o]$, $\text{TR}_{\text{SPGR}} = 6.8$, $\text{TR}_{\text{SSFP}} = 3.4$ & $\sum \text{TR}_i = 13.6 $ & $\Gamma_1 = 23.29, \Gamma_2 = 24.64$ \\ 
\hline 
SEIR & $\text{TR}_{\text{IR}} = 2994$, $\text{TI} = 1270$, $\text{TR}_{\text{SE}} = 2942$, $T_E = 17$ & TR = 7206 & $\Gamma_1 = 22.56$, $\Gamma_2 = 8.78$ \\ 
\hline 
LL & $\alpha = 30^o$, $t = [206:206:3090]$, N = 15, TR = 8900 & TR = 8900 & $\Gamma_1 = 21.32$ \\
\hline 
FIR1 & TI = [0 : 378 : 2268], W = 5647, N = 7 & $\sum \text{TR}_i = 47467 $ & $\Gamma_1 = 19.64$ \\
\hline 
FIR2 & TI = [0 : 303 : 2424], TR = 6722, N = 9 & $\sum \text{TR} = 60498 $ & $\Gamma_1 = 19.57$ \\
\hline  
CIR & TI = [0 : 450 : 1800], W = 10000, N = 5 & $\sum \text{TR}_i = 54500$ & $\Gamma_1 = 17.07$ \\ 
\hline 
SR & TI = [0 : 620 : 6820], N = 12 & $\sum \text{TI}_i = 40920$ & $\Gamma_1 = 7.52$  \\ 
\hline 
\end{tabular}
 
\end{center}
\vspace{1em}	
\label{table1}
\end{table*}

Table \ref{table1} shows the optimized sequence parameters for each relaxometry approach, along with the corresponding average $T_1$ and $T_2$ estimate efficiencies in the specified $T_1$ and $T_2$ ranges assuming the same input SNR level of $M_0/\sigma = 100$. The optimal pulse parameters of the CIR, SR, FIR, and LL sequences maximize the $T_1$ estimate efficiency. The optimal pulse parameters of the SEIR and DESPOT sequences maximize a weighted sum of the $T_1$ and $T_2$ estimates efficiencies. The DESPOT optimization uses the original acquisition scheme in \cite{despot} with two SPGR acquisitions and two SSFP acquisitions for $T_1$ and $T_2$ mapping. However for the selected $T_1$ and $T_2$ ranges, the DESPOT sequence optimization converges to the same two flip angles for the SPGR acquisitions. This surprising finding agrees with the DESPOT optimization results for adult brain in \cite{despotcrb2014}. To reduce this obvious redundancy, the optimization continues by using only one SPGR acquisition but doubling $\text{TR}_{\text{SPGR}}$. Doubling $\text{TR}_{\text{SPGR}}$ while keeping the $\text{TR}_{\text{SSFP}}$ the same will result in the same total scan time as the original DESPOT scan protocol in \cite{despot}. This new DESPOT sequence requires three acquisitions for each $T_1$/$T_2$ mapping (one SPGR and two SSFP), which is the minimum number of data acquisitions required for simultaneously estimating three unknowns $M_0$, $T_1$ and $T_2$. For the specific ranges of $T_1/T_2$ values corresponding to human brain white and grey matters, DESPOT has the highest $T_1$ estimate efficiency, followed by SEIR $>$ LL $>$ FIR1 $>$ FIR2 $>$ CIR $>$ SR. For $T_2$ estimation, DESPOT has a higher efficiency than SEIR. To validate the $T_1$ and $T_2$ estimate efficiencies predicted from the CRB, Monte Carlo trials were performed with results shown in the following section. \\

\subsection{Numerical Simulation Results}
\label{numericalsimlation}

This section presents the Monte Carlo simulation results to validate the $T_1$ and $T_2$ estimate efficiencies analysis and therefore confirm the effectiveness of CRB in predicting the $T_1$ and $T_2$ mapping performances. The parameter estimation processes involve joint $M_0$ and $T_1$ estimation for $T_1$ relaxometry approaches (CIR, SR, FIR and LL), and joint $M_0$, $T_1$ and $T_2$ estimation for $T_1$/$T_2$ relaxometry approaches (SEIR and DESPOT). For all relaxometry approaches, the simulated data from the signal models are generated by adding white Gaussian noise under different equivalent SNR levels shown in Table \ref{table2}. Equivalent SNR levels are calculated assuming a total scan time of 10 seconds for each relaxometry approach. Therefore, pulse sequences with a relatively short sequence time, such as DESPOT, can improve the input SNR through more signal averaging while pulse sequences with a relatively long sequence time, such as CIR, complete fewer sequences in the fixed scan time, reducing their SNR gain through signal averaging.

The NLSE minimization should not assume any prior knowledge of $M_0$, $T_1$ and $T_2$ values, such as the search boundaries, to avoid introducing biases to the estimate results. In this paper, the mean squared errors are minimized using an unconstrained nonlinear local optimization approach with the Nelder-Mead simplex direct search method \cite{Nelder}\cite{Optimization}. An initial value of $T_1$ = 2000 ms and $T_2$ = 200 ms are selected as the start point to initiate the simplex algorithm. For all simulations, a single expected value of $M_0 = 3000$ is assumed and five thousand independent Monte Carlo trials are repeated for each tested value of $T_1$ and $T_2$. Estimation results for all trials are recorded to calculate the variances and biases to compare against the CRB among all relaxometry approaches. 

Fig. 1 demonstrates the $T_1$ estimate precisions for the brain white and grey matters in terms of MEE (in symbols) for CIR, SR, FIR1, FIR2, LL, SEIR and DESPOT at equivalent input SNRs given in Table \ref{table2}. The dashed lines characterize the square root of the CRB on $T_1$ variance normalized by the true $T_1$ value, termed percentage CRB (PCRB). Table \ref{table2} shows the average MEE of $T_1$ estimate from Monte Carlo simulations for each $T_1$ relaxometry approach, compared against their theoretical PCRB. The MEE approaches closely their percentage CRB for the tested $T_1$ range. This implies the NLSE is the optimal estimator in the case of $T_1$ estimation, and also that the CRB provides a reliable prediction on the precision performance of each $T_1$ relaxometry approach. Fig. 1 shows for brain WM and GM with $T_1 \in$ [1000, 2000] ms DESPOT has the best precision for $T_1$ estimation, with $1.25\% \leq \text{MEE} \leq 1.52\%$. SEIR has precision close to DESPOT. LL has $1.48\% \leq \text{MEE} \leq 1.59\%$. The two types of FIR sequences have the same $T_1$ estimate precisions, with $1.69\% \leq \text{MEE} \leq 1.72\%$. CIR has $1.78\% \leq \text{MEE} \leq 1.98\%$. SR has the worst precision for $T_1$ estimation, with $4.18\% \leq \text{MEE} \leq 4.66\%$. Fig. 2 demonstrates the $T_1$ estimate accuracy in terms of relative biases, defined in (\ref{PBias}), for CIR, SR, FIR, LL, SEIR and DESPOT approaches. Overall for the tested $T_1$ range, the $T_1$ estimates are quite accurate with $-0.1\% \leq  \text{Rbias} \leq 0.2\%$. For most cases, the NLSE overestimates $T_1$. Table \ref{table2} shows the average relative biases within the $T_1$ range of interest. Among all approaches, FIR2 has the best accuracy with an average Rbias of $0.0005\%$, with the accuracies of FIR1, SEIR,  LL, DESPOT, CIR and SR decreasing accordingly.

\begin{table*}[!t]
\caption{$T_1$ and $T_2$ estimation performances for all relaxometry approaches. All equivalent SNR levels are calculated assuming a total scan time of 10 seconds for each relaxometry approach. The PCRB is calculated as the square root of CRB on $T_1$ or $T_2$ variance normalized over the true $T_1$ or $T_2$ value. The MEE and Rbias from Monte Carlo simulations are calculated following the definitions in (\ref{mse}) and (\ref{PBias}). }

\begin{center}
\begin{tabular}{|c|c|c|c|c|c|c|c|c|c|}
\hline 
$\textbf{Brain}$ & DESPOT ($T_1$) & SEIR ($T_1$) & LL & FIR1 & FIR2 & CIR & SR & DESPOT ($T_2$) & SEIR ($T_2$) \\ 
\hline 
Equi. MC SNR & 2711.63 & 117.80 & 105.99 & 45.89 & 40.66 & 42.84 & 49.43 & 2711.63 & 117.80\\ 
\hline 
PCRB $\%$ & 1.36 & 1.41 & 1.48 & 1.61 & 1.62 & 1.85 & 4.21 & 1.28 & 3.62\\ 
\hline 
MC MEE $\%$ & 1.37 & 1.41 & 1.49 & 1.62 & 1.62 & 1.86 & 4.24 & 1.28 & 3.63\\ 
\hline 
MC Rbias $\%$ & 0.012 & 0.008 & 0.009 & 0.006 & 0.0005 & 0.019 & 0.087 & 0.01 & 0.13\\ 
\hline 
\end{tabular} 
\end{center}
\label{table2}
\end{table*}
\begin{figure}
\begin{center}
\includegraphics[scale=0.21]{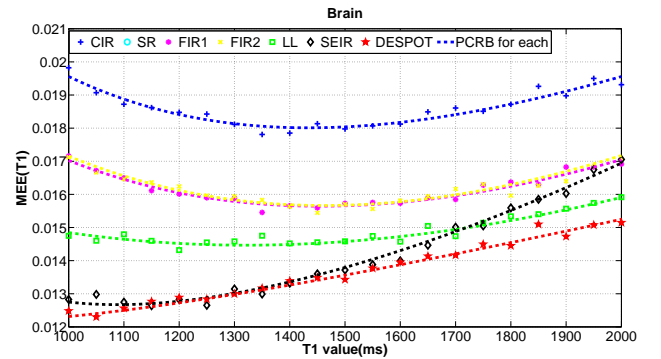}
\caption{Comparison on $T_1$ estimates' mean estimation error (MEE) for brain white and grey matters. Relaxometry approaches include CIR (blue), SR (cyan), FIR1 (magenta), FIR2 (yellow), LL (green), SEIR (black) and DESPOT (red). The dash lines characterize the theoretical percentage Cramer-Rao Bounds of unbiased $T_1$ estimation and the dots characterize simulation results for each approach. Note the SR approach is off the chart due to SR's relatively large MEE($T_1$) values compared to other approaches. All comparisons assume a total scan time of 10 seconds.}
\end{center}
\end{figure}

\begin{figure}
\begin{center}
\includegraphics[scale=0.21]{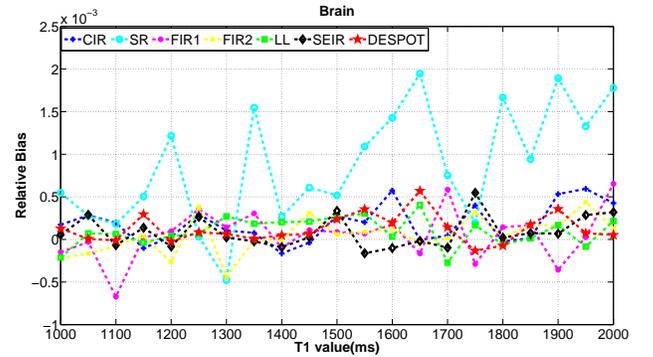}
\caption{Comparison on $T_1$ estimates' relative bias (Rbias) for brain white and grey matters for CIR (blue), SR (cyan), FIR1 (magenta), FIR2 (yellow), LL (green), SEIR (black) and DESPOT (red) approaches. All comparisons assume a total scan time of 10 seconds.}
\end{center}
\end{figure}

Fig. 3 demonstrates the $T_2$ estimate precision in terms of MEE (in symbols) for joint $T_1$/$T_2$ relaxometry approaches including SEIR and DESPOT at equivalent SNR levels given in Table \ref{table2}. The dashed lines in each panel characterize the percentage CRB of $T_2$ estimation for each relaxometry approach. Table \ref{table2} also shows the average MEE of $T_2$ estimation from Monte Carlo simulations for each $T_2$ relaxometry approach, compared against their theoretical PCRB. The MEE closely approaches the PCRB for the tested $T_2$ range, which implies the NLSE is the optimal estimator in the case of $T_2$ estimation, and also that the CRB provides a reliable prediction of the precision performance of each $T_2$ relaxometry approach. Fig. 3 shows for brain WM and GM with $T_2 \in$ [60, 110] ms, DESPOT has higher $T_2$ estimate precisions than SEIR with $1.24\% \leq \text{MEE} \leq 1.32\%$. SEIR has a worse $T_2$ estimate precision with $3.22\% \leq \text{MEE} \leq 4.11\%$. Fig. 4 demonstrates the $T_2$ estimate accuracy in terms of relative biases for SEIR and DESPOT approaches. Overall for the tested $T_2$ range, the $T_2$ estimates are quite accurate with $-0.05\% \leq \text{Rbias} \leq  0.25\%$. For most cases, the NLSE are over estimating $T_2$, where DESPOT has higher $T_2$ estimate accuracy with average Rbias of $0.01\%$ and SEIR has lower $T_2$ estimate accuracy with average Rbias of $0.13\%$.

\begin{figure}
\begin{center}
\includegraphics[scale=0.21]{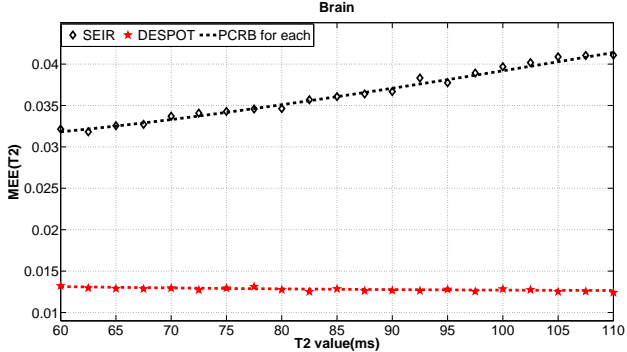}
\caption{Comparison on $T_2$ estimates' mean estimation error (MEE) for brain white and grey matters with SEIR (black) and DESPOT (red) approaches. The dash lines characterize the theoretical percentage Cramer-Rao Bounds of unbiased $T_2$ estimation and the dots characterize simulation results for each approach. All comparisons assume a total scan time of 10 seconds.}
\end{center}
\end{figure}

\begin{figure}
\begin{center}
\includegraphics[scale=0.21]{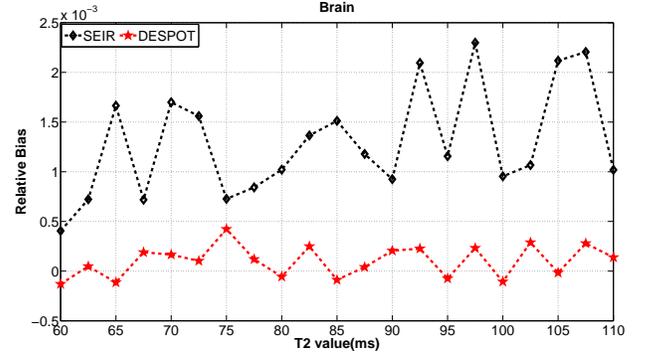}
\caption{Comparison on $T_2$ estimates' relative bias (Rbias) for brain white and grey matters with SEIR (black) and DESPOT (red) approaches. All comparisons assume a total scan time of 10 seconds.}
\end{center}
\end{figure}


\section{Discussion}
\subsection{Factors Controlling the TNR efficiency}
Eq. (\ref{efficiency}) establishes a common expression for $T_1$/$T_2$ estimation efficiencies for any relaxometry approach. Geometrically interpreting the CRB reveals that there are five factors affecting the TNR efficiency: the $T_1$/$T_2$ value, the sequence time $T_{\text{seq}}$, the input SNR, the sensitivity of the signal weighting vector $\textbf{h}$ to $T_1$/$T_2$, and the orthogonality between the signal weighting vector $\textbf{h}$ and the sensitivity vectors $\partial \textbf{h}/\partial T_{1}$ and $\partial \textbf{h}/\partial T_{2}$. These factors offer insight to improve the mapping efficiency and also to evaluate different relaxometry approaches. Among the factors, $T_1$ and $T_2$ are the parameters to be estimated, which largely depend on the tissue types and characteristics and therefore can't be controlled for improving efficiency. Decreasing the sequence time $T_{\text{seq}}$ helps improving efficiency since it increases the number of signal averaging within a fixed total scan time. Improving the  physical $\text{SNR} = M_0/\sigma$ (before any signal averaging) requires either increasing the $B_0$ field strength or employing less noisy receiver coil, which both increase the hardware costs. In contrast, increasing signal sensitivity and orthogonality improves $T_1$/$T_2$ relaxometry performance without incurring additional costs in hardware or scan time. Most previous sequence design research sought to improve the $T_1$/$T_2$ estimates precision by increasing the input SNR. Several prior studies \cite{despot}-\cite{oneshot} tried to improve the signals' dynamic ranges (DR), but failed to explicitly point out that DR is actually only one component of the signal sensitivity. No previous study recognized that improving the signals' orthogonality is also effective in improving the $T_1$/$T_2$ mapping efficiency. 

As an example, Figs. 5 and 6 compare the $T_1$ estimate sensitivity and orthogonality of different relaxometry approaches for brain WM and GM with $T_1 \in [1000, 2000]$ ms. Fig. 5 shows that overall the signal sensitivity decreases as $T_1$ increases. Among all sequences, FIR2 has the highest $T_1$ estimate sensitivity, with FIR1 and CIR sensitivities following closely. SEIR, LL and SR have similar and relatively lower $T_1$ estimate sensitivities than CIR. Surprisingly, DESPOT has the lowest $T_1$ estimate sensitivity. Fig. 6 shows that CIR, FIR1 and FIR2 have the highest $T_1$ estimate orthogonality between $[0.85, 1.00]$ for tested $T_1$ range. LL has orthogonality between $[0.78, 0.92]$. SEIR and SR have relatively lower orthogonality and DESPOT has the lowest orthogonality in $T_1$ estimation. These two panels explain the relatively high efficiencies of FIR1, FIR2 and LL in $T_1$ mapping due to their relatively high $T_1$ estimate sensitivity and orthogonality. The efficiency of CIR in $T_1$ mapping is not competitive largely due to its relatively long sequence time $T_{\text{seq}}$. SEIR has high $T_1$ estimate efficiency largely due to its relatively short sequence time and high orthogonality. SR has low $T_1$ estimate efficiency due to both its low sensitivity and orthogonality. DESPOT owes its high $T_1$ estimate efficiency much to the very short sequence time $T_{\text{seq}}$, which balances out DESPOT's poor sensitivity and orthogonality among all relaxometry approaches. For the tested $T_1$ and $T_2$ range, DESPOT has the highest $T_1$ and $T_2$ estimate efficiency among all relaxometry sequences largely due to its very short sequence time, but suffers from a relatively high Rbias even at very high SNR. This implies DESPOT has low accuracy issues with $T_1$ and $T_2$ mapping, which agrees with the findings in \cite{chang}\cite{MRF}. 

\begin{figure}
\begin{center}
\includegraphics[scale=0.21]{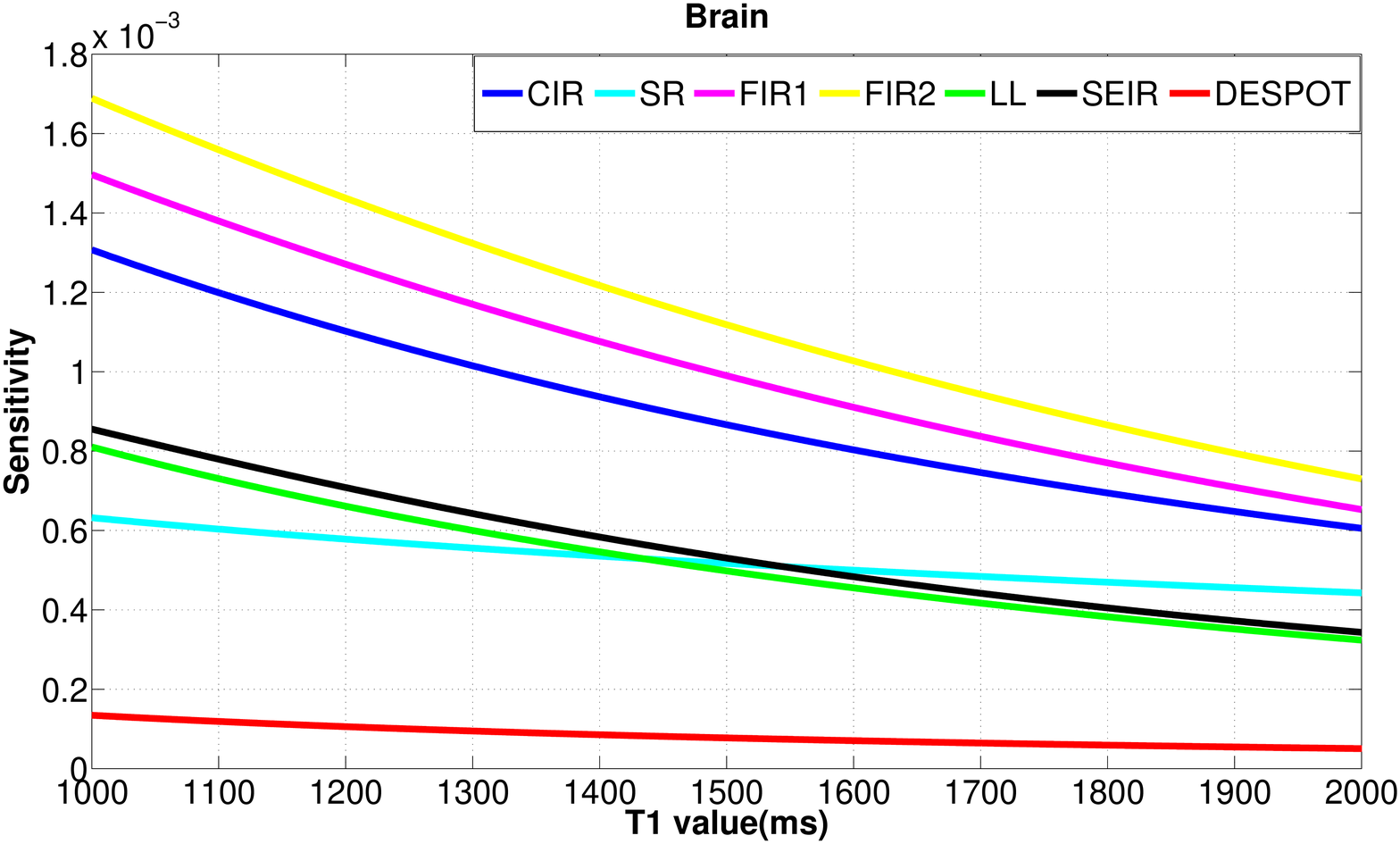}
\caption{Comparison on $T_1$ estimate sensitivity for brain WM and GM with CIR (blue), SR (cyan), FIR1 (magenta), FIR2 (yellow), LL (green), SEIR (black) and DESPOT (red) approaches.}
\end{center}
\end{figure}

\begin{figure}
\begin{center}
\includegraphics[scale=0.21]{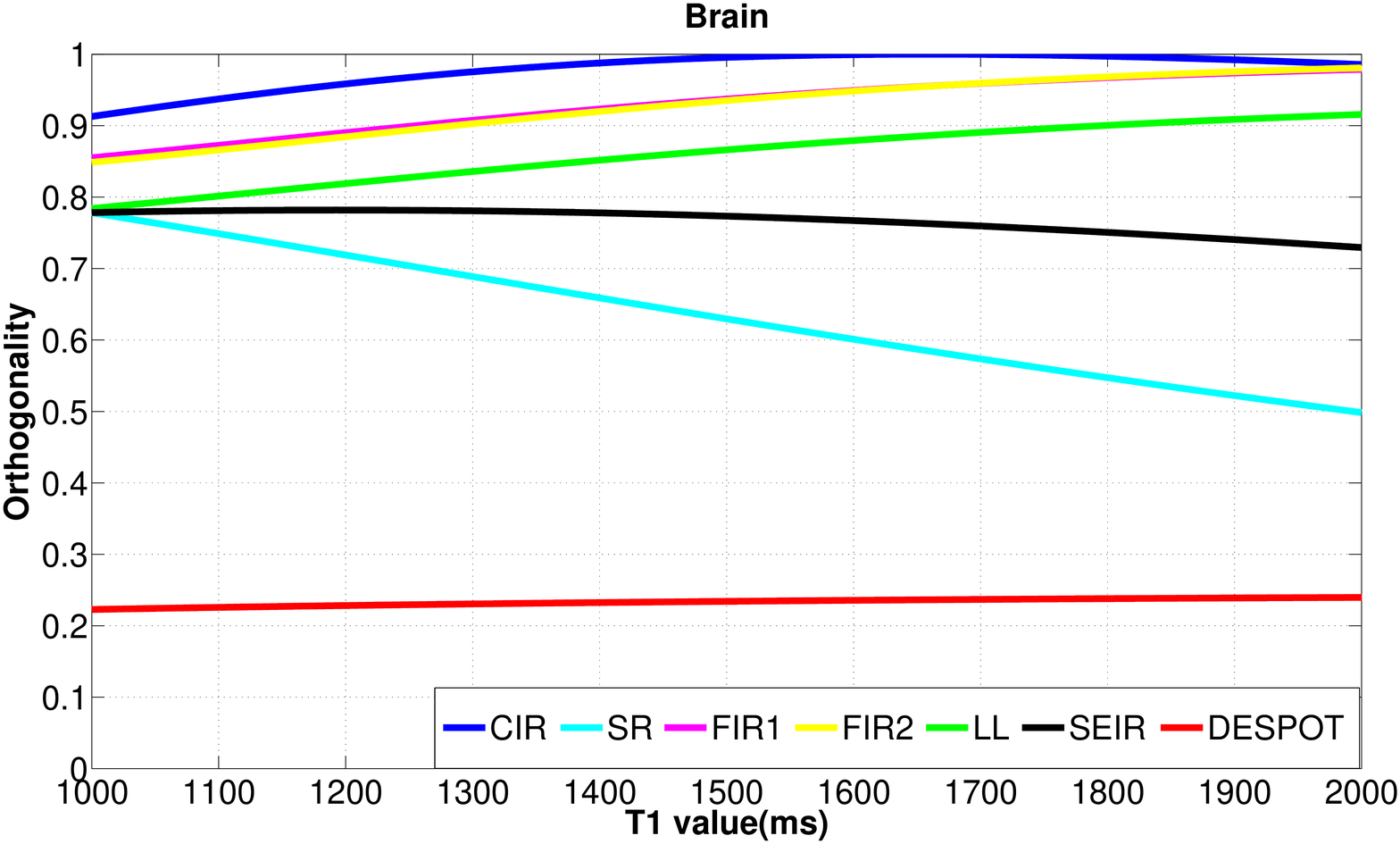}
\caption{Comparison on $T_1$ estimate orthogonality for brain WM and GM with CIR (blue), SR (cyan), FIR1 (magenta), FIR2 (yellow), LL (green), SEIR (black) and DESPOT (red) approaches.}
\end{center}
\end{figure}

\subsection{Practical Concerns}
Eq. (\ref{efficiency}) indicates that the TNR efficiency depends linearly on the input SNR. To ensure fair comparison, SNR equivalence between different relaxometry approaches needs to be calibrated by accounting for the physical scan parameters to match the simulated performances with practical experiments. The numerical simulations in the results section assume the use of the same MRI scanner to acquire data on the same physical object. Moreover, machine controllable scan parameters such as the receiver bandwidth, field-of-view (FOV) and voxel size are assumed to be the same across all relaxometry approaches to avoid unnecessary calibrations. However, the $T_2$ decay for spin echo acquisition sequences (IR, SR, SEIR sequences) and $T_2^*$ decay for gradient echo acquisition sequences (LL and DESPOT sequences) should not be ignored. For spin echo imaging, a longer echo time $T_E$ decreases the acquired MR signal amplitude and therefore decreases the input $\text{SNR}$ through $e^{-T_E/T_2}$. For gradient echo imaging, similarly, a longer echo time $T_E$ decreases the acquired MR signal amplitude through $e^{-T_E/T_2^*}$. For practical in vivo applications, both the $T_2$ and $T_E$ values for spin echo imaging are usually longer than the $T_2^*$ and $T_E$ values for gradient echo imaging. This results in comparable signal alternations for all sequences involved. Also, largely due to the incomplete knowledge of the average $T_2^*$ values for brain WM/GM, liver and blood, the $T_2$ and $T_2^*$ decays are therefore intentionally ignored in this paper. However, given a specific $T_E$ value for the data acquisition protocol and average $T_2$/$T_2^*$ values of the target tissue, this calibration could always be added in calculating the TNR efficiency. 

Another major source of error impacting the TNR efficiency in practical applications is the flip angle non-idealities, largely caused by patient-induced $B_1$ inhomogeneities due to the RF field distortions. Flip angle perturbations belong to systematic errors and can decrease the $T_1$ and $T_2$ mapping precisions and accuracies in the estimation process. Moreover, these errors can't be reduced through more signal averaging due to the inherent inaccurate signal modeling. Alternatively, quantitative $B_1$ mapping can be applied to calibrate the flip angles, but always at the cost of more scan time. The CRB derivations in this paper assume complete knowledge of $B_1$ inhomogeneities and thus the TNR efficiencies would be overestimated for experiments requiring extra scan time to calibrate $B_1$ inhomogeneities. Further work is required to explore quantitatively the effects of flip angle perturbations on the TNR efficiency, especially on the sensitivity and orthogonality factors for $T_1$ and $T_2$ estimation.

\section{Conclusion}
This paper establishes a statistical framework to optimize, evaluate and compare different relaxometry sequences based on their $T_1$/$T_2$ estimation characteristics. This framework considers relaxometry protocols as estimation algorithms and proposes two new metrics: the TNR to characterize precision and TNR efficiency to measure precision per unit time for $T_1$/$T_2$ estimators. The TNR and its efficiency are defined from the Cramer-Rao lower bound on the parameter estimate variance. Numerical results from Monte Carlo simulations achieve this statistical lower bound for all tested relaxometry approaches and thus validate the effectiveness of the CRB in predicting $T_1$/$T_2$ estimation performances. Geometrically the TNR efficiency definition reveals more completely the competing factors improving the $T_1$/$T_2$ mapping capabilities: increasing input SNR, decreasing sequence time, increasing signals' sensitivity to $T_1$/$T_2$ and also increasing the orthogonality between the signal vector and sensitivity vectors. This framework offers a reliable prediction on the $T_1$ and $T_2$ mapping performances of any relaxometry approaches before any phantom or in vivo experiments.

\end{document}